# *In situ* U–Pb chronology and chemistry of zirconolite in the andesitic meteorite Erg Chech 002

JUN SAKUMA[1], HISASHI ASANUMA[2,3], NAOTO TAKAHATA[2], AKIRA YAMAGUCHI[4,5], TSUYOSHI IIZUKA[1,*]

[1] Department of Earth and Planetary Science, The University of Tokyo, Bunkyo, Tokyo 113-0033, Japan

[2] Atmosphere and Ocean Research Institute, The University of Tokyo, Kashiwa, Chiba 277-0882, Japan

[3] Graduate School of Human and Environmental Studies, Kyoto University, Sakyo, Kyoto 606-8501, Japan

[4] National Institute of Polar Research, Tokyo 190-8518, Japan

[5] Department of Polar Science, School of Multidisciplinary Science, SOKENDAI, Tokyo 190-8518, Japan

*To whom correspondence should be addressed.

Email: iizuka@eps.s.u-tokyo.ac.jp

**ABSTRACT**

Precise and accurate ages for asteroidal crusts are fundamental for reconstructing the timeline of magmatic, metamorphic, and impact events in the early Solar System. Zirconolite ($CaZrTi_2O_7$) is an accessory mineral found in a wide range of crustal rocks on both the Earth and Moon, and has proven to be a potentially useful U–Pb chronometer. However, this mineral is rare in asteroidal meteorites, and its use for early Solar System chronology has been limited. We present the *in situ* occurrence, U–Pb chronology, and chemistry of zirconolite in the andesitic meteorite Erg Chech 002, which represents a sample of the oldest known asteroidal crust. The zirconolite occurs as needle- and fiber-shaped and stubby crystals with widths of ~3 µm and lengths of up to ~30 µm. Electron and ion microprobe analyses yielded concordant U–Pb data with a weighted mean $^{207}Pb/^{206}Pb$ age of 4557.9 ± 4.3 Ma (2σ), rendering it the Solar System's oldest known zirconolite. Yet, this age is distinctly younger than reported high-precision $^{207}Pb/^{206}Pb$ ages varying from 4565.6 to 4566.2 Ma, which were obtained by acid leaching of pyroxene and whole rock samples of the meteorite. From its mineralogical and REE–(U+Th)–(Nb+Ta) characteristics, we argue that the zirconolite age represents the timing of a shock metamorphism of the parent asteroid's crust. Our results suggest that $^{207}Pb/^{206}Pb$ dating for acid-leached samples can be affected by including even a tiny amount of metamorphic zirconolite, calling for caution in interpreting the high-precision $^{207}Pb/^{206}Pb$ age data. On the basis of thermodynamic and geochemical considerations, we infer further occurrences of zirconolite in alkali-silica-rich asteroidal rocks that rapidly cooled from high temperatures.

## INTRODUCTION

Zirconolite ($CaZrTi_2O_7$) is an accessory mineral found in a wide range of terrestrial rocks, including kimberlites, carbonatites, syenites, metamorphic rocks, and metasomatic rocks (Williams and Gieré, 1996; Gieré et al., 1998). It has also been detected in a variety of lunar rocks such as basalts, troctolites, KREEP-rich breccias, as well as in breccia clasts of anorthosite and granite (Frondel, 1975; Seddio et al., 2013). Since zirconolite's crystal structure can accommodate the substitution of elements with ionic radii from 0.40 Å to 1.14 Å and in valence states of 2+ to 6+ (Gieré et al., 1998), crystallizing zirconolite can incorporate large amounts of U (typically thousands of ppm) but little Pb (ionic radius of 1.19 Å), making it a potentially useful mineral for U–Pb chronology (Andersen and Hinthorne, 1972; Hinthorne et al., 1979; Rasmussen and Fletcher, 2004). Indeed, the precise timing of mare volcanism and basin-forming impacts on the Moon have been defined by $^{207}Pb/^{206}Pb$ ages of zirconolites from basalts and an anorthosite clast collected during the sample-return missions (Rasmussen et al., 2008; Norman and Nemchin, 2014; Li et al., 2021). In addition, many lunar and Martian meteorites contain minute crystals of zirconolite (Zeigler et al., 2005; Day et al., 2006; Wittmann et al., 2019; Hewins et al., 2020; Bechtold et al., 2021; Park and Kim, 2024), which have become an important target for *in situ* $^{207}Pb/^{206}Pb$ dating by high-spatial resolution ion microprobe (Zhang et al., 2010; Hao et al., 2021; Wang et al., 2021).

In contrast, there are only a few reports of zirconolite from asteroidal meteorites; to our knowledge, it has been documented only in the Allende CV chondrite (El Goresy et al., 1978; Sheng et al., 1991) and the silicate-bearing Miles IIE iron meteorite (Kirby et al., 2022). Micron-scale zirconolite grains occur in plagioclase-olivine-rich inclusions of the Allende chondrite, whereas relatively coarse zirconolite grains with sizes of ~10 μm occur in alkali-rich felsic silicate inclusions of the Miles iron meteorite. The relatively large Miles zirconolite grains enabled *in situ* $^{207}Pb/^{206}Pb$ dating by ion microprobe, yielding a reasonably precise $^{207}Pb/^{206}Pb$ age of 4539 ± 12 Ma (all errors are 2σ) (Kirby et al., 2022). Yet, ubiquitousness, formation conditions, and significance of asteroidal zirconolite remain to be further investigated.

Herein we report the occurrence of zirconolite with sizes up to 30 μm in the unique meteorite Erg Chech 002 (EC 002). This meteorite has an andesite bulk composition and consists mainly of plagioclase, pyroxene, and silica minerals, with pyroxene and olivine xenocrysts reaching several centimeters (Barrat et al., 2021; Jin et al., 2024). The igneous texture and elemental zoning patterns of an olivine xenocryst in EC 002 indicate its rapid crystallization at or near the surface of the parent asteroid (Barrat et al., 2021; Beros et al., 2024). The major and trace element chemistry suggest that EC 002 formed by partial melting of a metal-free chondritic precursor with no depletion in alkali elements (Barrat et al., 2021; Nicklas et al., 2022). The meteorite has been dated with several radiometric systems, including short-lived $^{26}$Al–$^{26}$Mg (Barrat et al., 2021; Fang et al., 2022; Connelly et al., 2023; Reger et al., 2023), $^{53}$Mn–$^{53}$Cr (Zhu et al., 2022; Anand et al., 2022; Yang et al., 2025), $^{60}$Fe–$^{60}$Ni (Fang et al., 2025), and $^{146}$Sm–$^{142}$Nd (Fang et al., 2022) systems and long-lived $^{40}$K–$^{40}$Ar (Barrat et al., 2021), $^{40}$K–$^{40}$Ca (Dai et al., 2023), $^{147}$Sm–$^{143}$Nd (Fang et al., 2022), and $^{238,235}$U–$^{206,207}$Pb systems (Connelly et al., 2023; Krestianinov et al., 2023; Reger et al., 2023). These results show that EC 002 crystallized within a few Myrs after the formation of Ca–Al-rich inclusions (CAIs), making it a piece of the oldest known asteroidal crust. In detail, however, the ages obtained using the same radiometric systems by different studies are inconsistent with each other. The $^{207}$Pb/$^{206}$Pb age of 4565.56 ± 0.12 Ma obtained by acid leaching for pyroxene as well as whole rock samples by Krestianinov et al. (2023) is distinctly younger than that of 4566.19 ± 0.20 Ma for whole rock samples by Connelly et al. (2023). The reported $^{26}$Al–$^{26}$Mg and $^{53}$Mn–$^{53}$Cr ages also span over ~0.5 Myr and ~2 Myr, exceeding the analytical uncertainties of ~0.01 Myr and ~0.6 Myr, respectively. These age discrepancies may be attributed to 1) isotopic disturbance during a secondary thermal event such as shock metamorphism, 2) initial isotopic heterogeneity within the samples, as inferred from the presence of xenocrysts, and/or 3) analytical artifacts (Fang et al., 2022; Zhu et al., 2022; Anand et al., 2022; Amelin et al., 2025; Yang et al., 2025). Solving the age controversy is central for understanding the thermal history of the parent asteroid (Neumann et al., 2023) and for building consistent chronology in the early Solar

System. The latter is because the high-precision isotope data for EC 002 can potentially be used to anchor the short-lived radiometric chronometers onto the absolute $^{207}$Pb/$^{206}$Pb timescale. While CAIs and quenched angrites are widely used as time anchors, establishing a new reliable time anchor is crucial for investigating the distribution of the short-lived radionuclides in the protosolar disk and their stellar origins (e.g., Iizuka et al., 2025). Considering the presence of xenocrysts in EC 002, *in situ* $^{207}$Pb/$^{206}$Pb dating of a U-rich mineral is a critical approach to address the controversy. Reger et al. (2023) reported an ion microprobe $^{207}$Pb/$^{206}$Pb age of 4564.3 ± 5.2 Ma for merrillite. In this study, we present *in situ* $^{207}$Pb/$^{206}$Pb dating and chemistry of zirconolite in EC 002.

## METHODS

We prepared three polished thin sections of a 7.3 g stone of EC 002. To investigate the detailed texture and to identify the Zr-bearing as well as phosphate minerals, elemental maps (Al, Ca, Cr, Fe, Mg, Na, P, Si, Ti, and Zr) were obtained for the thin sections using a JEOL JXA-8530F field emission-electron probe microanalyzer (FE-EPMA) at the Department of Earth and Planetary Science, University of Tokyo. Analytical conditions for the elemental mapping consisted of a 15 kV acceleration voltage, 80 nA beam current, 10 μm beam diameter, and 10 μm step size for the stage motion. The chemical compositions of the identified zirconolite grains were further determined using the FE-EPMA with a 20 kV acceleration voltage, 70 nA beam current, and 1 μm beam diameter. The standard materials used for the measurements were as follows (standards expressed by chemical formulas are synthetic crystals): albite (Na); MgO (Mg); $Al_2O_3$ (Al); $CaSiO_3$ (Si and Ca); $CuFeS_2$ (S); adularia (K); $KTiPO_4$ (P); $TiO_2$ (Ti and Mn); chromite (Cr); hematite (Fe); NiO (Ni); $SrBaNb_4O_{12}$ (Sr and Nb); $Y_3Al_5O_{12}$ (Y); $ZrO_2$ (Zr); synthetic rare earth element (REE) phosphate (lanthanoids); $HfO_2$ (Hf); tantalum metal (Ta); tungsten metal (W); PbVGe (Pb); thorianite (Th); and uraninite (U). The backgrounds and spectral interferences were corrected following the methods of Williams et al. (1995) and Ichimura et al. (2020). The peak intensities were then corrected for ZAF (atomic number-absorption-fluorescence) effects.

The Pb isotope compositions of the zirconolite grains were determined using a Cameca NanoSIMS 50 ion microprobe at the Atmosphere and Ocean Research Institute, University of Tokyo. Details of the instrument design are given in Takahata et al. (2008). An 0.1–0.2 nA mass-filtered primary O$^-$ beam was focused onto an ~2 μm diameter spot on the sample surface that was coated with gold. Ion images with $^{48}$Ti$^+$ and $^{90}$Zr$^+$ on the target mineral areas were used to precisely locate analytical spots. An array of electron multipliers, in dynamic mode, was set to monitor the following ions: $^{28}$Si$^+$ (EM1), $^{48}$Ti$^+$ (EM2), $^{90}$Zr$^+$ (EM3), $^{204}$Pb$^+$ (EM4), $^{206}$Pb$^+$ (EM4B), $^{238}$U$^{16}$O$^+$ (EM5), and $^{238}$U$^{16}$O$_2^+$ (LD) in the first line, and $^{207}$Pb$^+$ (EM4B) in the second line. The entrance and exist slits were adjusted to obtain a mass resolving power of ~9,000 at the shoulder with 10–90% peak height, which is sufficient to resolve Pb$^+$ from problematic molecular interferences (e.g., Zr$_2$O$^+$). Data were acquired over 30 cycles with 2 lines/cycle, 10 s integration/line, and 3 s idle time between lines. The relative yields of the electron multipliers were calibrated against NIST SRM 610 glass. Instrumental mass bias for Pb isotopes was assumed to be negligible on the basis of results from previous Pb isotope analyses by NanoSIMS (Stern et al., 2005; Sano et al., 2006; Yang et al., 2012). No zirconolite standard is currently available for U abundance determination or Pb/U fractionation correction; therefore, the NanoSIMS Pb/U data are provided in Online Materials[1] Table DR1, but not included in the following discussion.

Determination of REE abundances in the zirconolite grains was also carried out on the NanoSIMS. The settings of the primary beam and slits were identical to those used in the Pb isotope analysis. The array of four electron multipliers (EM2–5) was set to measure all lanthanides with dynamic mode following the protocol described in Shi et al. (2022), whereas EM1 was used to monitor $^{90}$Zr$^+$. Data were obtained over 50 cycles with 6 lines/cycle, 3 s integration/line, and 3 s idle time between lines. The total counts for each REE$^+$ were referenced to $^{90}$Zr$^+$ and the ZrO$_2$ concentration in zirconolite. The relative sensitivity factors of REE$^+$/$^{90}$Zr$^+$ were determined using NIST SRM 610 (Pearce et al., 1997), given that the matrix effect between silicate glass and Zr-mineral is sufficiently small (Shi et al., 2022).

**RESULTS**

Five zirconolite grains were identified in the three thin sections (Fig. 1a-e), along with the other Zr-minerals baddeleyite (Fig. 1e) and zircon (Fig. 1f). Baddeleyite is considerably more abundant than zirconolite and zircon. Two of the zirconolite grains, Zrc#1 and Zrc#2 (Fig. 1a,b), were relatively large with lengths of ~30 μm and widths of ~3 μm and had needle- and fiber-like shapes, respectively. They occurred along grain boundaries of albitic plagioclase and pyroxene in association with silica and Fe metal. The other three zirconolite grains were a few μm in size and had low aspect ratios. Grain Zrc#3 (Fig. 1c) also occurred at a boundary between plagioclase and pyroxene, while grain Zrc#4 (Fig. 1d) was located at a plagioclase-silica boundary associated with merrillite, troilite, and an unidentified Ti-Fe-Zr mineral. Grain Zrc#5 (Fig. 1e) was surrounded by plagioclase, together with globules of ilmenite, baddeleyite, merrillite, and an unidentified Nb-Ti-Zr mineral.

The FE-EPMA data from nine spots on the five zirconolite grains are summarized in Table 1. The EC 002 zirconolite was significantly depleted in the major components (3.92–8.57 wt% CaO, 0.34–8.42 wt% $ZrO_2$, and 10.49–27.43 wt% $TiO_2$) compared to the ideal zirconolite composition (CaO = 16.54 wt%, $ZrO_2$, = 36.34 wt%, and $TiO_2$ = 47.12 wt%). In a complementary fashion, it contained substantial amounts of MgO, $Al_2O_3$, $SiO_2$, $Cr_2O_3$, MnO, FeO, $Y_2O_3$, $Nb_2O_5$, $HfO_2$, $Ta_2O_5$, PbO, $ThO_2$, and $UO_2$. The contents of $ThO_2$ and $UO_2$ ranged from 0.60 to 6.86 wt% and from 0.17 to 2.11 wt%, respectively. The contents of $Na_2O$, $SO_3$, $K_2O$, NiO, SrO, $REE_2O_3$, and $WO_3$ were under or close to the detection limits (~0.10–0.15 wt%). Compositional variations among different grains were more significant than those within single grains. In particular, Zrc#5 was markedly enriched in FeO, $Nb_2O_5$, PbO, $ThO_2$, and $UO_2$ compared to the other grains.

The NanoSIMS REE data from three spots on the two large zirconolite grains are shown in Table 2 and plotted in a chondrite-normalized diagram together with those from terrestrial carbonatites, lunar basalts and noritic anorthosite (Fig. 2). Both EC 002 zirconolite grains showed slight LREE depletion and moderately negative Eu anomalies (Eu/Eu* = 0.12–0.21). The EC 002 and lunar zirconolite grains were similar in the degree of LREE depletion, but differed in the Eu anomaly level and total REE abundance. The combined NanoSIMS and FE-EPMA data averaged for individual zirconolite grains are plotted in a triangular diagram of ΣREE–(U+Th)–(Nb+Ta) (Fig. 3). For grains Zrc#3–5 that were not analyzed by NanoSIMS for REEs, their ΣREE/Y values were assumed to be identical to the average of grains Zrc#1 and Zrc#2. When compared with compositional variations of previously studied zirconolite grains of various origins, the EC 002 zirconolite was most similar to terrestrial metamorphic zirconolite.

The NanoSIMS Pb isotope data from fourteen spots on four zirconolite grains are reported in Table 3. All zirconolite grains had highly radiogenic Pb isotope compositions with $^{204}Pb/^{206}Pb$ of <0.0001. We calculated radiogenic $^{207}Pb/^{206}Pb$ ratios ($^{207}Pb^*/^{206}Pb^*$) assuming that the non-radiogenic Pb component had the primordial Pb isotope composition of Blichert-Toft et al. (2010). The $^{207}Pb^*/^{206}Pb^*$ calculations are unaffected by the choice of the primordial or terrestrial common Pb isotope composition due to the low levels of non-radiogenic Pb. Analyses of fourteen spots yielded a weighted average $^{207}Pb^*/^{206}Pb^*$ of 0.6212 ± 0.0019 (2σ, MSWD = 1.8) (Fig. 4), which corresponds to a date of 4557.9 ± 4.3 Ma using the $^{238}U/^{235}U$ value of 137.8288 ± 0.0054 (2σ) reported for the EC 002 whole-rock samples (Krestianinov et al., 2023). Due to the extremely radiogenic Pb isotope composition of the EC 002 zirconolite, the U/Pb and Th/Pb elemental ratios can be regarded as a function of the age. These elemental ratios calculated from the FE-EPMA data (Table 1) are plotted along the 4558 Ma reference isochron (Fig. 5), indicating that the U-Th-Pb system has remained closed since approximately that time.

## DISCUSSION

**Occurrence of zirconolite in asteroidal meteorites**

All five zirconolite grains found in EC 002 occur along with albitic plagioclase, and three of them are closely associated with silica (Fig. 1). Similar close associations of zirconolite with alkali-feldspar and silica have been observed in the Miles iron meteorites and lunar rocks (Rasmussen et al., 2008; Kirby et al., 2022). However, the co-existence of zirconolite and silica is rather unusual in terrestrial rocks: terrestrial zirconolite commonly occurs in silica-undersaturated rocks such as kimberlites and syenites (Williams and Gieré, 1996; Gieré et al., 1998). Moreover, terrestrial metamorphic rocks contain evidence that zirconolite was destabilized by addition of $SiO_2$ (Pan et al., 1997; Urueña et al., 2023). To account for the mineral assemblage of zircon + titantie + rutile formed after zirconolite (Pan et al., 1997) and for the crystalline phase stability in glass-ceramic wasteforms (Maddrell et al., 2015), the following reaction has been proposed:

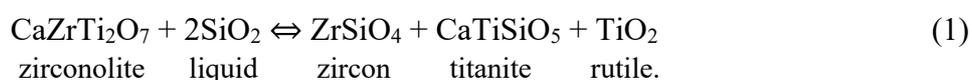

$$CaZrTi_2O_7 + 2SiO_2 \Leftrightarrow ZrSiO_4 + CaTiSiO_5 + TiO_2 \qquad (1)$$
zirconolite    liquid    zircon    titanite    rutile.

Taking into account the Gibbs free energies of formation of these minerals at room temperature, Maddrell et al. (2015) further suggested that zirconolite is thermodynamically stable only at low silica activity. To better understand the relation between the silica activity and zirconolite stability, we have calculated the silica activity for a melt in equilibrium with pure zirconolite, zircon, titanite and rutile at high temperatures. For the standard state of $SiO_2$ in a melt, silica glass at the temperature of interest has been used. We have also calculated the silica activity for a melt in equilibrium with a pure silica mineral for comparison. The results (Fig. 6) indicate that pure zirconolite can crystallize from a silica-oversaturated magma at temperatures higher than ~1200 °C. Writing reaction (1) in terms of silica minerals instead of melt, it is also indicated that pure zirconolite cannot stably coexist with silica minerals below ~1200 °C. Note, however, that the concentration of non-ideal components in natural zirconolite can be significantly high (Williams and Gieré, 1996; Gieré et al., 1998) (Table 1). This non-ideality substantially decreases the activity of $CaZrTi_2O_7$ in zirconolite and, in turn, lowers the temperature limit for the co-existence of zirconolite and silica during magmatism and metamorphism.

In light of the thermodynamic stability, the close association of zirconolite and silica particularly in extra-terrestrial samples may reflect that extra-terrestrial magmas are generally volatile-poor and tend to crystallize at higher temperatures compared to terrestrial magmas. For instance, the bulk composition of Ca-rich pyroxene in EC 002 indicates a crystallization temperature of ~1200 ˚C (Barrat et al., 2021). In addition, transient heating by impacts on asteroids could be favorable for secondary zirconolite formation even in silica-oversaturated rocks. By contrast, the co-existence of zirconolite and silica should be prohibited by thermal metamorphism with slow cooling. Indeed, zircon rather than zirconolite crystallized at 900–950 ˚C during prolonged thermal metamorphism in basaltic eucrites, the most abundant group of achondrites (Iizuka et al., 2015; Iizuka et al., 2019; Barboni et al., 2024).

Besides the thermal condition, the zirconolite occurrence is controlled by the chemistry of magmas/rocks. The interstitial occurrence of zirconolite-silica-alkali feldspar assemblage in extra-terrestrial samples indicates that these minerals, if of magmatic origin, precipitated from residual melts in the late-stage of crystallization. For zirconolite crystallization, Zr saturation in the melts should be critical and likely occurred as a result of both the increase of melt Zr concentration through crystallization of Zr-poor phases and the decrease of Zr solubility in the melts with elevating $SiO_2$ contents (Crisp and Berry, 2022). In this respect, it is noteworthy that neither zirconolite nor other Zr-minerals have been found from angrites—a group of basaltic achondrites that are ultra-depleted in alkali elements and enriched in refractory elements such as Ti and Zr—except for baddeleyite from the anomalous angrite Angra dos Reis (Prinz et al., 1977; Keil, 2002). Specifically, bulk Zr concentrations of typical angrites range from 34 to 70 ppm (Warren et al., 1995; Mittlefehldt et al., 2002; Kurat et al., 2004), which are a factor of 2 to 4 higher than that of EC 002 (Barrat et al., 2021), and overlap well with the range of terrestrial kimberlites (Giuliani et al., 2025; Zech et al., 2025). The lack of Zr-minerals in angrites may be partly attributed to ultrabasic compositions ($SiO_2$ = 33–43wt%; Keil, 2012) leading to high Zr solubility of the melts from which they crystallized. However, this cannot explain the widespread occurrence of Zr-minerals in kimberlites having even lower $SiO_2$ contents (Scatena-Wachel and Jones 1984; Williams and Gieré, 1996; Gieré et al., 1998). Rather, Zr saturation in angrite melts was probably prevented by strong partitioning of Zr into Al-Ti-bearing pyroxene (traditionally called fassaite), as indicated by its high Zr concentrations of ~100 to >300 ppm (Floss et al., 2003). Such Al-Ti-bearing pyroxene appears as a major liquidus phase characteristically in alkali-free systems (Longhi, 1999). Overall, we infer that the potential of zirconolite occurrence is relatively high in asteroidal meteorites that are rich not only in Zr but also in silica and alkali elements, especially if they rapidly cooled from temperatures as high as ~1200 ˚C.

**Significance of the zirconolite age**

The EC 002 zirconolite $^{207}$Pb/$^{206}$Pb age of 4557.9 ± 4.3 Ma pushes back the oldest known zirconolite by ~20 Myr. The zirconolite age overlaps within uncertainty with the reported merrillite $^{207}$Pb/$^{206}$Pb age of 4564.3 ± 5.2 Ma (Reger et al., 2023), but it is distinctly younger than the reported $^{207}$Pb/$^{206}$Pb ages by acid leaching for pyroxene and whole rock samples, ranging from 4565.56 ± 0.12 Ma to 4566.19 ± 0.20 Ma (Reger et al., 2023; Connelly et al., 2023; Krestianinov et al., 2023). All the $^{26}$Al–$^{26}$Mg and $^{53}$Mn–$^{53}$Cr isochrons for EC 002 also yield absolute ages older than 4565 Ma, irrespective of whether anchored to CAIs or the D'Orbigny angrite (Barrat et al., 2021; Fang et al., 2022; Zhu et al., 2022; Anand et al., 2022; Connelly et al., 2023; Reger et al., 2023; Yang et al., 2025). We consider four alternative scenarios to explain the younger zirconolite $^{207}$Pb/$^{206}$Pb age in relation to its mineralogical and chemical characteristics.

First, considering that EC 002 contains xenocrysts (Barrat et al., 2021; Zhu et al., 2022), analysis of a mixture of phenocrysts and xenocrysts could result in an apparent older crystallization age. It may be possible therefore that the zirconolite $^{207}$Pb/$^{206}$Pb age of ~4558 Ma reflects the timing of crystallization, whereas the older ages are biased due to mixed analyses. However, this scenario requires that xenocrysts are dominant over phenocrysts within the analyzed samples dated as ~4566 Ma, unless the xenocrysts have presolar ages (i.e., >4567 Ma), which is extremely unlikely. Furthermore, because no plagioclase xenocryst has been identified in EC 002, the $^{26}$Al–$^{26}$Mg isochron age determined by *in situ* plagioclase analysis (Barrat et al., 2021) should not be the subject of this issue, precluding this scenario.

The second scenario is that the U–Pb system of the zirconolite was selectively reset *via* diffusion during a secondary thermal event at 4558 Ma. Although Pb diffusion in zirconolite has not been quantified experimentally, Wu et al. (2010) theoretically estimated Pb closure temperatures of zirconolite based on the kinetic-porosity model of Zhao and Zheng (2007), and reported closure temperatures ranging from ~600 to ~900 ˚C for a 1 μm radius zirconolite grain at a cooling rate of 100 ˚C/Myr. The closure temperature depended on whether the polytype was two-layered monoclinic (~700 ˚C), three-layered orthorhombic (~600 ˚C), or three-layered trigonal (~900 ˚C). The polytype of the EC 002 zirconolite could not be determined by our Raman analysis which yielded a featureless spectrum reflecting accumulation of α recoil damage. Note that the cumulative α dose in the EC 002 zirconolite during the first 10 Myr after crystallization was below the critical value of $4 \times 10^{18}$ α/g at which this mineral becomes amorphous at ambient temperature (Deschanels et al., 2014). On the other hand, augite with a ~1 mm radius is expected to have a far higher Pb closure temperature (~1500 ˚C) at a similar cooling rate, based on the diffusion parameters of Cherniak (2001). Hence, a secondary thermal event could cause preferential resetting of the zirconolite U–Pb system over the pyroxene U–Pb system. However, this scenario is difficult to reconcile with the old $^{26}$Al–$^{26}$Mg isochron ages. Since the $^{26}$Al–$^{26}$Mg isochron is defined mainly by plagioclase with a ~1 mm radius, its closure temperature can be approximated to be ~600 ˚C (Van Orman et al., 2014), which is comparable to the lowest predicted closure temperature of the zirconolite U–Pb system. In addition, while one zirconolite grain analyzed in this study (Zrc #5) has markedly higher U and Th contents than the other grains (Table 1), they have identical $^{207}$Pb/$^{206}$Pb ages (Fig. 4). This is opposed to lunar zirconolite whose $^{207}$Pb/$^{206}$Pb ages were variably reset depending on U-Th contents during an impact event (Norman and Nemchin, 2014).

The third scenario is that the EC 002 zirconolite was formed during a secondary thermal event at 4558 Ma, whereas the major constituent minerals primarily crystallized from andesitic magma at ~4565 Ma. We favor this scenario for the following three reasons. 1) Metamorphic zirconolite can be formed by a reaction of Zr-bearing phases with Ca- and Ti-bearing phases (Purtscheller and Tessadri, 1985; Hornig and Wörner, 1991; Tropper et al., 2007). The co-existence of the zirconolite grain Zrc#5 with baddeleyite, ilmenite, and merrillite (Fig. 1) is suggestive of such metamorphic reaction. 2) The ΣREE–(Th+U)–(Nb+Ta) relative abundance of the EC 002 zirconolite is comparable to that of terrestrial metamorphic zirconolite. 3) The mottled optical extinction of plagioclase and pyroxenes and deformation in augite indicate that EC 002 underwent shock metamorphism (Barrat et al., 2021; Beros et al., 2024). On the other hand, the EC 002 zirconolite shows no diagnostic features of shock metamorphism such as polycrystalline textures and metallic Pb nano-inclusions (Zhang et al., 2024). Accordingly, we interpret the younger zirconolite $^{207}$Pb/$^{206}$Pb age as reflecting the timing of zirconolite growth during this shock metamorphism. If this interpretation is correct, the metamorphic temperature should be high enough so that zirconolite could be formed in the presence of silica (Fig. 6).

The last but not the least possibility is an artifact caused by instrumental mass fractionation (IMF) during Pb isotope analysis. Previous studies with NanoSIMS (Stern et al., 2005; Sano et al., 2006; Yang et al., 2012) have demonstrated that IMF is negligible when a single collector is used for measuring Pb isotopes with magnetic peak switching. Our analyses of NIST SRM 610 glass and NBS SRM 981 Pb metal also yielded non-IMF-corrected $^{207}$Pb/$^{206}$Pb ratios of $0.9216 \pm 0.0153$ and $0.9135 \pm 0.0015$ (2σ), which are consistent with the reference values of 0.9098 (Woodhead and Hergt, 2001) and 0.9147 (Woodhead et al., 1995), respectively. Accordingly, we have applied no IMF correction for Pb isotope ratio measurements of the EC 002 zirconolite. Strictly speaking, however, IMF for Pb could substantially depend on the sample matrix. Although Stern et al. (2005) showed that the IMF was undetectable for zirconolite at the precision of 0.75%, a –0.5% $^{207}$Pb/$^{206}$Pb shift can account for the discrepancy from the short-lived radiometric ages. A rigorous test of this possible analytical artifact requires analyzing a zirconolite with a known Pb isotope composition that can be measured by NanoSIMS with a precision of better than 0.5%, but no such zirconolite is currently available.

IMPLICATIONS

This study provides new insights into the controversy surrounding the age of the andesitic meteorite EC 002. We argue that the zirconolite $^{207}Pb/^{206}Pb$ age of 4557.9 ± 4.3 Ma presents the timing of shock metamorphism, even though the possibility of uncorrected instrumental mass bias cannot be excluded. If this argument is correct, our results imply a significant influence of metamorphic zirconolite on $^{207}Pb/^{206}Pb$ age determinations from magmatic minerals such as pyroxene. The high-precision but variable $^{207}Pb/^{206}Pb$ isochron ages of EC 002 were obtained by acid leaching for pyroxene and whole rock samples with masses of ~20 to 400 mg (Reger et al., 2023; Connelly et al., 2023; Krestianinov et al., 2023). Considering that zirconolite is commonly adjacent to pyroxene (Fig. 1), zirconolite could have been included in the pyroxene samples, not to mention the whole rock samples. Since zirconolite has an orders-of-magnitude higher U concentration (~0.5wt%; Table 1) than the whole rock (~250 ppb) and pyroxenes (~600 ppb) (Krestianinov et al., 2023), even a tiny proportion of zirconolite can cause a detectable change in a $^{207}Pb/^{206}Pb$ age. For instance, assuming a magmatic crystallization age of 4566 Ma for EC 002, inclusions of 1 and 10 μg/g of 4558 Ma metamorphic zirconolite within a whole rock sample result in –0.16 and –1.33 Ma changes, respectively. We propose that the observed $^{207}Pb/^{206}Pb$ age variation among the pyroxene and whole rock samples can be attributed to variable proportions of included zirconolite. If so, even the oldest reported $^{207}Pb/^{206}Pb$ age of 4566.19 ± 0.20 Ma (Connelly et al., 2023) could represent a minimum estimate of the magmatic age. Actually, the oldest $^{207}Pb/^{206}Pb$ age was obtained from the whole rock samples rather than the pyroxene samples, although the former are expected to include zirconolite at higher abundances than the latter. This apparent contradiction may arise from different protocols of acid leaching: the old $^{207}Pb/^{206}Pb$ age is defined mainly by leachates of 1 M and 7 M HF from the whole rock samples, whereas the young age is obtained primarily from residues after $HNO_3$ and HCl leaching of the pyroxene as well as whole rock samples. Given that zirconolite is highly resistant to acid leaching (Nikoloski et al., 2019), it would be concentrated in the residues rather than the HF leachates, leading to a more prominent downward age shift in the residues. Besides this zirconolite effect, HF leaching could potentially cause $^{207}$Pb/$^{206}$Pb age biases through (i) fractionation of radiogenic Pb isotopes due to the size difference between α-recoil tracks of the $^{238}$U and $^{235}$U decay chains (Amelin et al., 2025) and (ii) differential dissolution between high-Ca and low-Ca lamellae in pyroxene that underwent exsolution during a secondary thermal event (Ito et al., 2019).

On the basis of the zirconolite occurrence in the andesitic meteorite EC 002 as well as other extra-terrestrial samples, this study suggests that zirconolite would be rather ubiquitous in alkali-silica-rich and rapidly cooled rocks. A recent experimental study (Collinet and Grove, 2022) indicates that such rocks would be widespread in small bodies of the early Solar System. In fact, an increasing number of andesitic meteorites have been found over the last two decades (Day et al., 2009; Bischoff et al., 2014; Srinivasan et al., 2018; Barrat et al., 2021). Hence, we anticipate discovering more asteroidal zirconolite, which will be an important target of *in situ* U–Pb dating in combination with mineralogy and chemistry. To fulfill the potential of zirconolite as an *in-situ* U–Pb chronometer for the early Solar System, however, a necessary step is establishing a zirconolite standard to rigorously evaluate the potential Pb isotopic mass bias as well as to correct for Pb/U fractionation during analysis.

## ACKNOWLEDGEMENTS AND FUNDING

We are grateful to H. Yoshida and K. Ichimura for analytical supports, to T. Mikouchi, H. Hayashi, and S. Yamazaki for assistance with sample preparation, and to Y. Amelin for discussion. Constructive reviews by A. Treiman and M. Norman improved the manuscript, and the editorial handling by D. Harlov and P. Tomascak is acknowledged. This work is financed by the Japan Society for the Promotion of Science (Grants #19H01959, #21KK0057, and #22H00170).

Bischoff, A., Horstmann, M., Barrat, J.-A., Chaussidon, M., Pack, A., Herwartz, D., Ward, D., Vollmer, C., and Decker, S., 2014. Trachyandesitic volcanism in the early Solar System. Proceedings of the National Academy of Sciences of the United States of America, 111, 12689–12692, https://doi.org/10.1073/pnas.1404799111.

Blichert-Toft, J., Zanda, B., Ebel, D.S., and Albarède, F. (2010) The Solar System primordial lead. Earth and Planetary Science Letters, 300, 152–163, https://doi.org/10.1016/j.epsl.2010.10.001.

Cherniak, D. J. (2001) Pb diffusion in Cr diopside, augite, and enstatite, and consideration of the dependence of cation diffusion in pyroxene on oxygen fugacity. Chemical Geology, 177, 381–397, https://doi.org/10.1016/S0009-2541(00)00421-6.

Cocherie, A. and Albarede, F. (2001) An improved U-Th-Pb age calculation for electron microprobe dating of monazite. Geochimica et Cosmochimica Acta, 65, 4509–4522, https://doi.org/10.1016/S0016-7037(01)00753-0.

Collinet, M. and Grove, T.L. (2020) Widespread production of silica- and alkali-rich melts at the onset of planetesimal melting. Geochimica et Cosmochimica Acta, 277, 334–357, https://doi.org/10.1016/j.gca.2020.03.005.

Connelly, J.N., Bollard, J., Amsellem, E., Schiller, M., Larsen, K.K., and Bizzarro, M. (2023) Evidence for very early planetesimal formation and $^{26}$Al/$^{27}$Al heterogeneity in the protoplanetary disk. The Astrophysical Journal Letters, 952, L33, https://doi.org/10.3847/2041-8213/ace42e.

Crisp, L.J. and Berry, A.J. (2022) A new model for zircon saturation in silicate melts. Contributions to Mineralogy and Petrology, 177, 71, https://doi.org/10.1007/s00410-022-01925-6.

Dai, W., Moynier, F., Fang, L., and Siebert, J. (2023) K–Ca dating and Ca isotope composition of the oldest solar system lava, Erg Chech 002. Geochemical Perspective Letters, 24, 33–37, https://doi.org/10.7185/geochemlet.2302.
17

achondrite? Geochimica et Cosmochimica Acta, 68, 1901–1921, https://doi.org/10.1016/j.gca.2003.10.016.

Li, Q.-L., Zhou, Q., Liu, Y., Xiao, Z., Lin, Y., Li, J.-H., Ma, H.-X., Tang, G.-Q., Guo, S., Tang, X., Yuan, J.-Y., Li, J., Wu, F.-Y., Ouyang, Z., Li, C., and Li, X.-H. (2021) Two-billion-year-old volcanism on the Moon from Chang'e-5 basalts. Nature, 600, 54–58, https://doi.org/10.1038/s41586-021-04100-2.

Longhi, J. (1999) Phase equilibrium constraints on angrite petrogenesis. Geochimica et Cosmochimica Acta, 63, 573–585, https://doi.org/10.1016/S0016-7037(98)00280-4.

Maddrell, E., Thornber, S., and Hyatt, N.C. (2015) The influence of glass composition on crystalline phase stability in glass-ceramic wasteforms. Journal of Nuclear Materials, 456, 461–466, https://doi.org/10.1016/j.jnucmat.2014.10.010.

McDonough, W.F. and Sun, S.-s. (1995) The composition of the Earth. Chemical Geology, 120, 223–253, https://doi.org/10.1016/0009-2541(94)00140-4.

Mittlefehldt, D.W., Killgore, M., and Lee, M.T. (2002) Petrology and geochemistry of D'Orbigny, geochemistry of Sahara 99555, and the origin of angrites. Meteoritics and Planetary Science, 37, 345–369, https://doi.org/10.1111/j.1945-5100.2002.tb00821.x.

Neumann, W., Luther, R., Trieloff, M., Reger, P.M., and Bouvier, A. (2023) Fitting thermal evolution models to the chronological record of Erg Chech 002 and modeling the ejection conditions of the meteorite. The Planetary Science Journal, 4, 196, https://doi.org/10.3847/PSJ/acf465.

Nicklas, R.W., Day, J.M.D., Gardner-Vandy, K.G., and Udry, A. (2022) Early silicic magmatism on a differentiated asteroid. Nature Geoscience, 15, 696–699, https://doi.org/10.1038/s41561-022-00996-1.

Nikoloski, A.N., Gilligan, R., Squire, J., and Maddrell, E.R. (2019) Chemical stability of zirconolite for proliferation resistance under conditions typically required for the leaching of highly refractory uranium minerals. Metals, 9, 1070, https://doi.org/10.3390/met9101070.
21

in mare basalt 10047. Geochimica et Cosmochimica Acta, 72, 5799–5818, https://doi.org/10.1016/j.gca.2008.09.010.

Reger, P.M., Roebbert, Y., Neumann, W., Gannoun, A., Regelous, M., Schwarz, W.H., Ludwig, T., Trieloff, M., Weyer, S., and Bouvier, A. (2023) Al–Mg and U–Pb chronological records of Erg Chech 002 ungrouped achondrite meteorite. Geochimica et Cosmochimica Acta, 343, 33–48, https://doi.org/10.1016/j.gca.2022.12.025.

Robie, R.A. and Hemingway, B.S. (1995) Thermodynamic properties of minerals and related substances at 298.15 K and 1 Bar ($10^5$ Pascals) pressure and at higher temperatures. U.S. Geological Survey Bulletin, 2131, 461. U.S. Department of the Interior.

Sano, Y., Takahata, N., Tsutsumi, Y., and Miyamoto, T. (2006) Ion microprobe U-Pb dating of monazite with about five micrometer spatial resolution. Geochemical Journal, 40, 597–608, https://doi.org/10.2343/geochemj.40.597.

Scatena-Wachel, D.E. and Jones, A.P. (1984) Primary baddeleyite ($ZrO_2$) in kimberlite from Benfontein, South Africa. Mineralogical Magazine, 48, 257–26, https://doi.org/10.1180/minmag.1984.048.347.10.

Seddio, S.M., Jolliff, B.L., Korotev, R.L., and Zeigler, R.A., 2013. Petrology and geochemistry of lunar granite 12032,366-19 and implications for lunar granite petrogenesis. American Mineralogist, 98, 1697–1713, https://doi.org/10.2138/am.2013.4330.

Sheng, Y.J., Hutcheon, I.D., and Wasserburg, G.J. (1991) Origin of plagioclase-olivine inclusions in carbonaceous chondrites. Geochimica et Cosmochimica Acta 55, 581–599, https://doi.org/10.1016/0016-7037(91)90014-V.

Shi, L., Sano, Y., Takahata, N., Koike, M., Morita, T., Koyama, Y., Kagoshima, T., Li, Y., Xu, S., and Liu, C. (2022) NanoSIMS analysis of rare earth elements in silicate glass and zircon: Implications for partition coefficients. Frontiers in Chemistry, 10, 844953, https://doi.org/10.3389/fchem.2022.844953.
23

Williams, C.T., Jones, A.P., and Wall, F. (1995) Analysis of rare earth minerals. In A.P. Jones, F. Wall, and C.T. Williams, Eds., Rare Earth Minerals: Chemistry, Origin, and Ore Deposits, Mineralogical Society Series, 7, p. 327–348. Chapman and Hall, England.

Williams, C.T. and Gieré, R. (1996) Zirconolite: A review of localities worldwide, and a compilation of its chemical compositions. Bulletin of the Natural History Museum London, 52, 1–24.

Wittmann, A., Korotev, R.L., Jolliff, B.L., and Carpenter, P.K. (2019) Spinel assemblages in lunar meteorites Graves Nunataks 06157 and Dhofar 1528: Implications for impact melting and equilibration in the Moon's upper mantle. Meteoritics and Planetary Science, 54, 379–394, https://doi.org/10.1111/maps.13217.

Woodhead, J.D., Volker, F., and McCulloch, M.T. (1995) Routine lead isotope determinations using a lead-207–lead-204 double spike: a long-term assessment of analytical precision and accuracy. Analyst, 120, 35–39, https://doi.org/10.1039/AN9952000035.

Woodhead, J.D. and Hergt, J.M. (2001) Strontium, neodymium and lead isotope analyses of NIST glass certified reference materials: SRM 610, 612, 614. Geostandards Newsletter, 25, 261–266, https://doi.org/10.1111/j.1751-908X.2001.tb00601.x.

Wu, F.-Y., Yang, Y.-H., Mitchell, R.H., Bellatreccia, F., Li, Q.-L., and Zhao, Z.-F. (2010) In situ U–Pb and Nd–Hf–(Sr) isotopic investigations of zirconolite and calzirtite. Chemical Geology, 277, 178–195, https://doi.org/10.1016/j.chemgeo.2010.08.007.

Yang, B., Pang, R., Wang, Q., Zhang, A., Du, W., and Qin, L. (2025) Highly heterogeneous parent body of the rare andesitic Erg Chech 002 meteorite revealed by the revisited Mn–Cr isotopic systematics. The Planetary Science Journal, 6, 73, https://doi.org/10.3847/PSJ/ada769

Yang, W., Lin, Y.-T., Zhang, J.-C., Hao, J.-L., Shen, W.-J., and Hu, S., 2012. Precise micrometre-sized Pb-Pb and U-Pb dating with NanoSIMS. Journal of Analytical Atomic Spectrometry, 27, 479–487, https://doi.org/10.1039/C2JA10303F.

**Figure captions**

**Figure 1:** Back-scattered electron images showing zirconolite grains (a) Zrc#1, (b) Zrc#2, (c) Zrc#3, (d) Zrc#4, (e) Zrc #5, and (f) a zircon grain in Erg Chech 002. Pits in Zrc#1 and Zrc#2 are the *in-situ* Pb isotope analysis spots from NanoSIMS. Phase abbreviations are as follow: Ap, apatite; Bdy, baddeleyite; Ilm, ilmenite; Mer, merrillite; Mt, metal; Pl, plagioclase; Px, pyroxene; Si, unidentified silica polymorph; Trd, tridymite; Tro, troilite; X, unidentified Ti-Fe-Zr mineral; Y, unidentified Nb-Ti-Zr mineral; Zrc, zirconolite; Zrn, zircon.

**Figure 2**: Chondrite-normalized REE patterns of zirconolite grains Zrc#1 and Zrc#2 in Erg Chech 002. The patterns in terrestrial carbonatites (Wu et al., 2010), lunar basalts (Rasmussen et al., 2008), and lunar noritic anorthosite (Norman and Nemchin, 2014) are shown for comparison. The chondrite data of McDonough and Sun (1995) were used for the normalization.

**Figure 3:** Triangular plot of ΣREE–(Th+U)–(Nb+Ta) for the Erg Chech 002 zirconolite. Also shown for comparison are zirconolite compositional fields for lunar rocks and terrestrial syenites, carbonatites, skarn, (ultra-)mafic rocks, metamorphic rocks, and metasomatic rocks. Data and source references are given in Online Materials Table DR2.

**Figure 4:** Radiogenic $^{207}Pb/^{206}Pb$ ratios and the weighted average (bold line) with 95% confidence interval (shaded area). Box heights are 2σ.

**Figure 5:** Plot of Th/Pb versus U/Pb for the Erg Chech 002 zirconolite. Errors at 95% confidence level are represented by ellipses. Also shown are chemical reference isochrons of 4558, 4400, 4200, and 4000 Ma that were obtained following Cocherie and Albarede (2001).

**Figure 6:** Variation of silica activity ($a_{SiO_2}$) with temperature at 1 bar for two silica buffer reactions: $SiO_2$ (glass) ⇔ $SiO_2$ (quartz/cristobalite) and $CaZrTi_2O_7$ (zirconolite) + $2SiO_2$ (glass) ⇔ $ZrSiO_4$ (zircon) + $CaTiSiO_5$ (titanite) + $TiO_2$ (rutile). Thermodynamic data are taken from Putnum et al. (1999) for zirconolite and from Robie and Hemingway (1995) for the other phases.

Table 1
Electron probe microanalysis data for the Erg Chech 002 zirconolite.

| Grain | Zrl#1 | Zrl#1 | Zrl#1 | Zrl#2 | Zrl#2 | Zrl#2 | Zrl#3 | Zrl#4 | Zrl#5 |
|---|---|---|---|---|---|---|---|---|---|
| Spot | EPMA-1 | EPMA-2 | EPMA-3 | EPMA-1 | EPMA-2 | EPMA-3 | EPMA-1 | EPMA-1 | EPMA-1 |
| MgO | 0.17 | 0.17 | 0.17 | 0.20 | 0.18 | 0.30 | 0.13 | 0.13 | 0.24 |
| $Al_2O_3$ | 0.40 | 0.19 | 0.31 | 0.16 | 0.18 | 0.18 | 1.13 | 1.06 | 0.36 |
| $SiO_2$ | 2.41 | 0.56 | 7.67 | 0.30 | 0.36 | 1.05 | 5.90 | 8.62 | 1.01 |
| $P_2O_5$ | b.d.l. | b.d.l. | b.d.l. | b.d.l. | b.d.l. | b.d.l. | b.d.l. | b.d.l. | 0.20 |
| CaO | 11.69 | 11.83 | 11.03 | 11.91 | 11.89 | 12.27 | 12.62 | 12.45 | 7.97 |
| $TiO_2$ | 33.98 | 33.94 | 31.84 | 34.92 | 35.00 | 34.51 | 36.64 | 31.43 | 19.69 |
| $Cr_2O_3$ | 0.48 | 0.51 | 0.45 | 0.57 | 0.57 | 0.56 | 0.47 | 0.57 | 0.55 |
| MnO | 0.31 | 0.35 | 0.31 | 0.28 | 0.34 | 0.35 | 0.25 | 0.30 | 0.34 |
| FeO[a] | 5.72 | 5.63 | 6.17 | 5.35 | 5.61 | 5.75 | 3.84 | 5.00 | 9.93 |
| $Y_2O_3$ | 0.17 | 0.19 | 0.18 | 0.32 | 0.31 | 0.32 | 0.33 | 0.08 | 0.68 |
| $ZrO_2$ | 33.94 | 36.00 | 33.35 | 34.38 | 34.43 | 35.46 | 33.48 | 30.59 | 27.92 |
| $Nb_2O_5$ | 6.70 | 6.76 | 6.31 | 6.29 | 6.19 | 6.08 | 5.02 | 10.16 | 15.44 |
| $HfO_2$ | 0.48 | 0.51 | 0.49 | 0.57 | 0.57 | 0.61 | 0.46 | 0.57 | 0.36 |
| $Ta_2O_5$ | 0.26 | 0.27 | 0.28 | 0.37 | 0.31 | 0.39 | 0.23 | 0.44 | 0.95 |
| PbO | 0.63 | 0.63 | 0.62 | 0.80 | 0.81 | 0.74 | 0.40 | 0.39 | 4.32 |
| $ThO_2$ | 0.96 | 0.97 | 0.96 | 1.05 | 1.08 | 1.00 | 0.62 | 0.60 | 6.86 |
| $UO_2$ | 0.31 | 0.31 | 0.29 | 0.44 | 0.44 | 0.38 | 0.18 | 0.17 | 2.11 |
| Total | 98.60 | 98.82 | 100.42 | 97.92 | 98.25 | 99.94 | 101.69 | 102.57 | 98.91 |
| | | | | | | | | | |
| Th/Pb | 1.457 | 1.451 | 1.474 | 1.242 | 1.257 | 1.278 | 1.472 | 1.460 | 1.502 |
| Th/Pb %error | 2.8% | 2.8% | 2.8% | 2.8% | 2.8% | 2.8% | 3.0% | 3.1% | 2.8% |
| U/Pb | 0.463 | 0.458 | 0.438 | 0.528 | 0.513 | 0.486 | 0.418 | 0.418 | 0.463 |
| U/Pb %error | 7.1% | 7.1% | 7.6% | 5.1% | 5.2% | 5.9% | 12.0% | 12.4% | 2.8% |
| Error-correl. | 20% | 20% | 19% | 28% | 27% | 24% | 12% | 12% | 50% |

All concentrations are presented in wt%.
[a] All iron was calculated as FeO considering the close association of zirconolite with Fe metal.

Table 2

Ion microprobe REE data for the Erg Chech 002 zirconolite.

| Grain | Zrl#1 | Zrl#1 | Zrl#2 |
|---|---|---|---|
| Spot | REE-1 | REE-2 | REE-1 |
| La | 83 | 113 | 131 |
| Ce | 348 | 464 | 629 |
| Pr | 63 | 94 | 144 |
| Nd | 342 | 493 | 938 |
| Sm | 137 | 204 | 471 |
| Eu | 9 | 11 | 15 |
| Gd | 130 | 141 | 299 |
| Tb | 21 | 27 | 42 |
| Dy | 238 | 273 | 555 |
| Ho | 60 | 63 | 132 |
| Er | 142 | 172 | 333 |
| Tm | 28 | 36 | 70 |
| Yb | 218 | 327 | 664 |
| Lu | 23 | 25 | 39 |
|  |  |  |  |
| $[La/Sm]_N$ | 0.38 | 0.35 | 0.17 |
| Eu/Eu* | 0.19 | 0.21 | 0.12 |

Note: All concentrations are presetend in ppm.

Table 3

Ion microprobe Pb isotope data for the Erg Chech 002 zirconolite.

| Grain | Spot | $^{204}Pb/^{206}Pb$ | 2SE | $^{207}Pb/^{206}Pb$ | 2SE | $^{207}Pb*/^{206}Pb*$ | error | Age | error |
|---|---|---|---|---|---|---|---|---|---|
| Zrl#1 | Pb-1 | 0.000030 | 0.000034 | 0.6163 | 0.0056 | 0.6162 | 0.0056 | 4546 | 13 |
| Zrl#1 | Pb-2 | 0.000014 | 0.000022 | 0.6224 | 0.0130 | 0.6223 | 0.0130 | 4560 | 30 |
| Zrl#1 | Pb-3 | 0.000044 | 0.000066 | 0.6282 | 0.0104 | 0.6280 | 0.0104 | 4573 | 24 |
| Zrl#1 | Pb-4 | 0.000095 | 0.000087 | 0.6250 | 0.0133 | 0.6246 | 0.0134 | 4566 | 31 |
| Zrl#1 | Pb-5 | 0.000014 | 0.000008 | 0.6261 | 0.0034 | 0.6260 | 0.0034 | 4569 | 8 |
| Zrl#1 | Pb-6 | 0.000027 | 0.000023 | 0.6200 | 0.0047 | 0.6199 | 0.0047 | 4555 | 11 |
| Zrl#2 | Pb-1 | 0.000017 | 0.000014 | 0.6230 | 0.0041 | 0.6229 | 0.0041 | 4562 | 10 |
| Zrl#2 | Pb-2 | 0.000010 | 0.000012 | 0.6158 | 0.0054 | 0.6157 | 0.0054 | 4545 | 13 |
| Zrl#2 | Pb-3 | 0.000019 | 0.000011 | 0.6185 | 0.0049 | 0.6184 | 0.0049 | 4551 | 11 |
| Zrl#2 | Pb-4 | 0.000003 | 0.000008 | 0.6268 | 0.0098 | 0.6267 | 0.0098 | 4571 | 22 |
| Zrl#2 | Pb-5 | 0.000029 | 0.000009 | 0.6189 | 0.0036 | 0.6187 | 0.0036 | 4552 | 8 |
| Zrl#2 | Pb-6 | 0.000029 | 0.000019 | 0.6223 | 0.0056 | 0.6222 | 0.0056 | 4560 | 13 |
| Zrl#3 | Pb-1 | 0.000073 | 0.000028 | 0.6200 | 0.0046 | 0.6196 | 0.0046 | 4554 | 11 |
| Zrl#5 | Pb-1 | 0.000012 | 0.000005 | 0.6221 | 0.0027 | 0.6221 | 0.0027 | 4560 | 6 |

\* Radiogenic Pb, corrected for using $^{204}Pb$ and the primordial Pb isotope composition of Blichert-Toft et al. (2010).

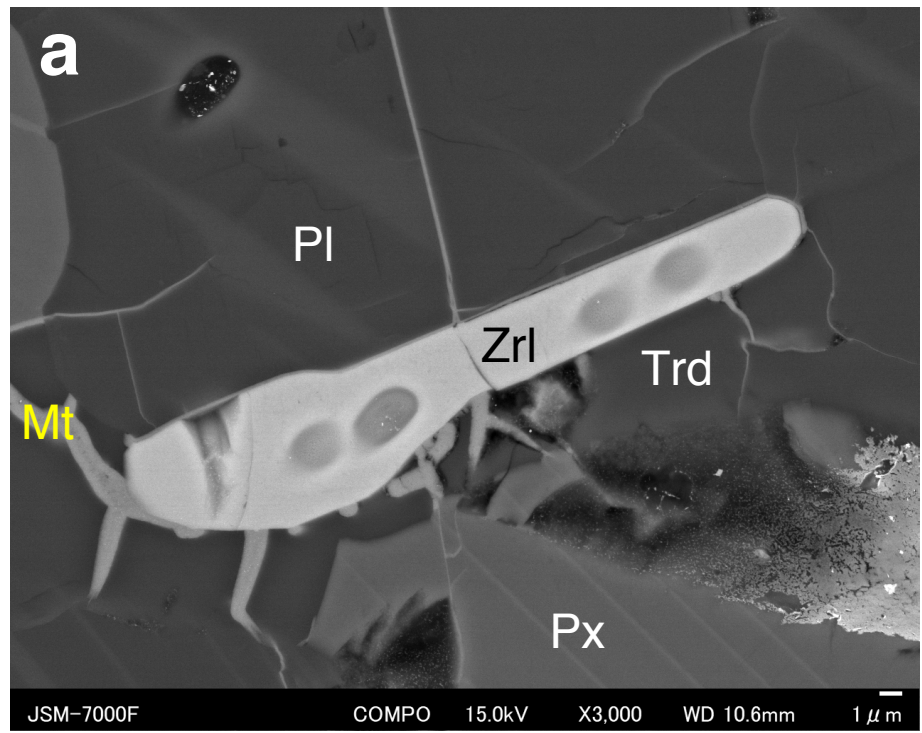
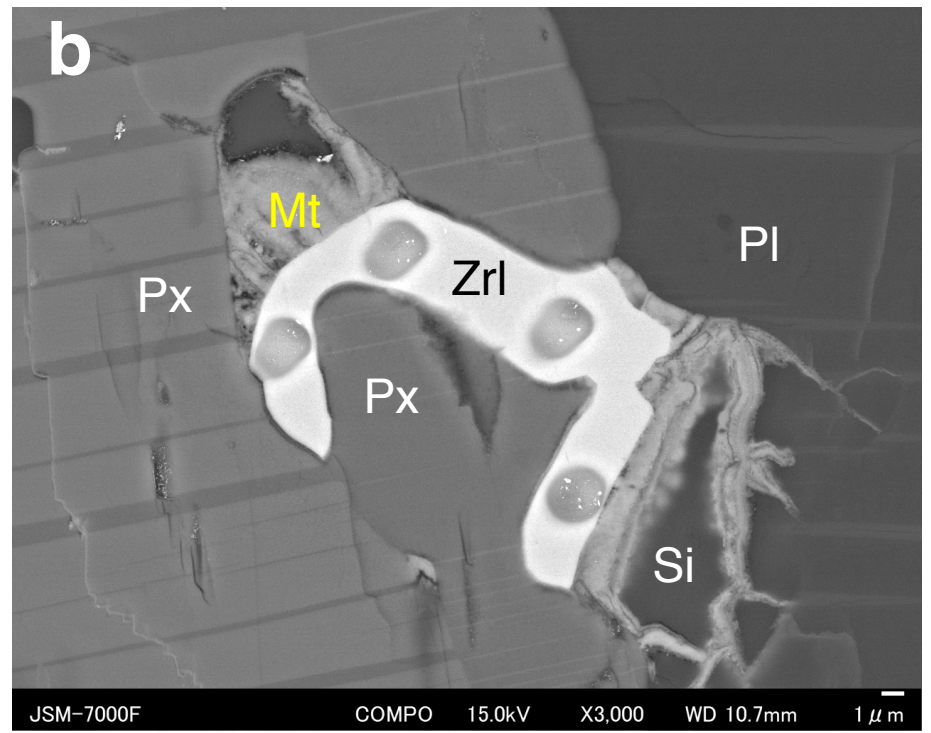
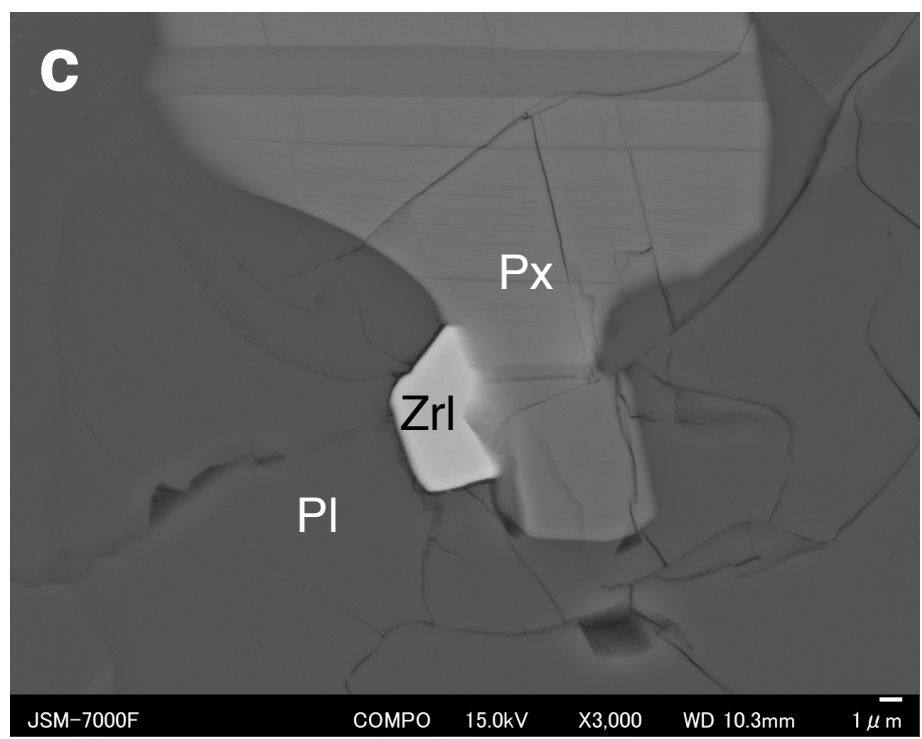
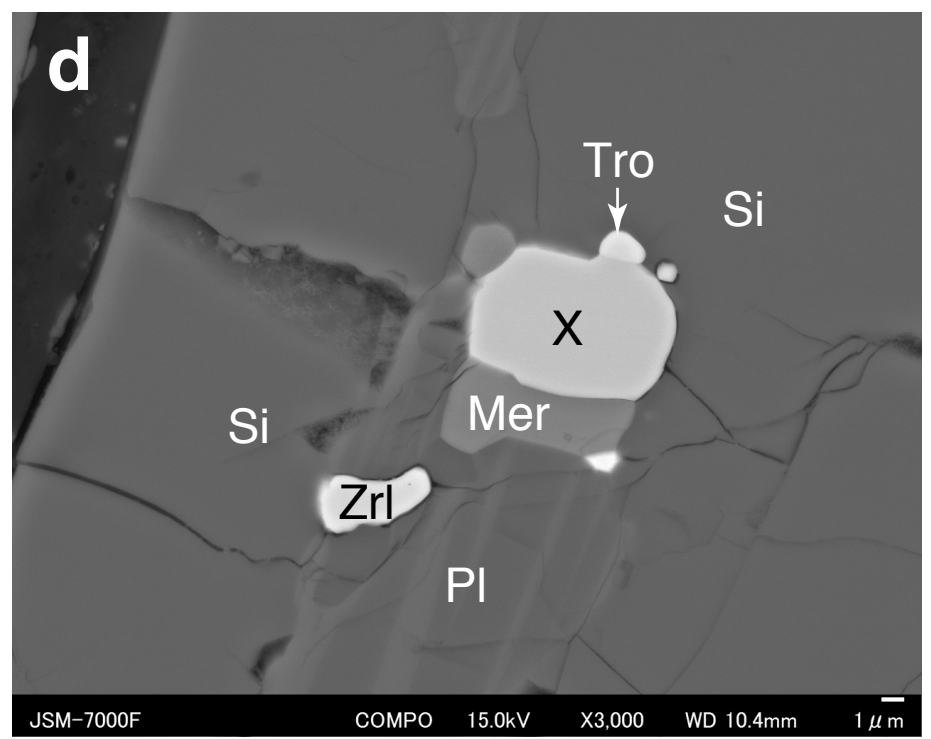
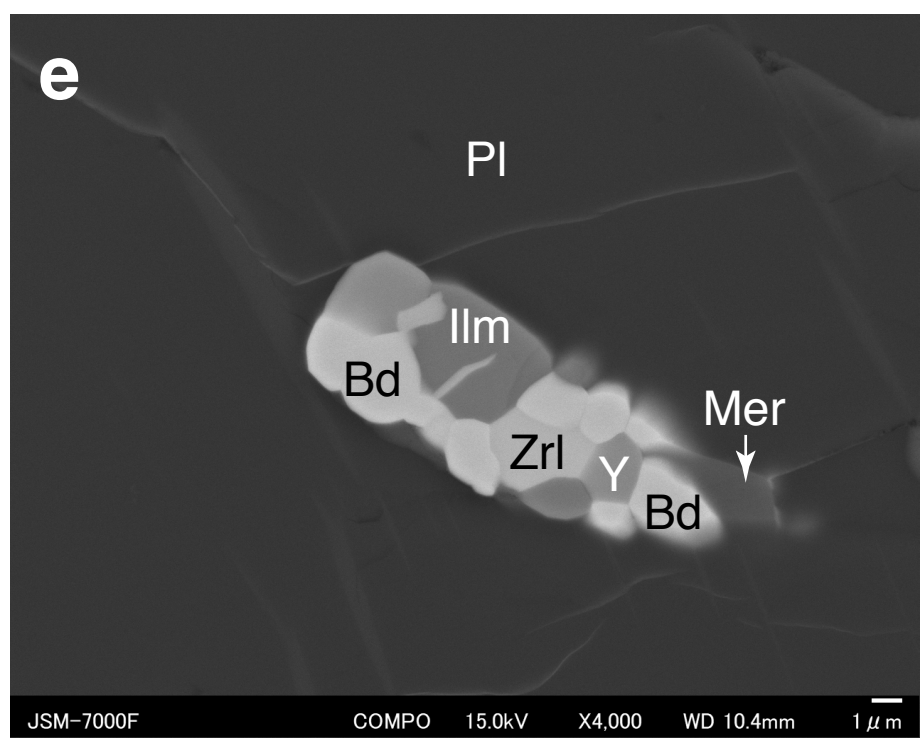
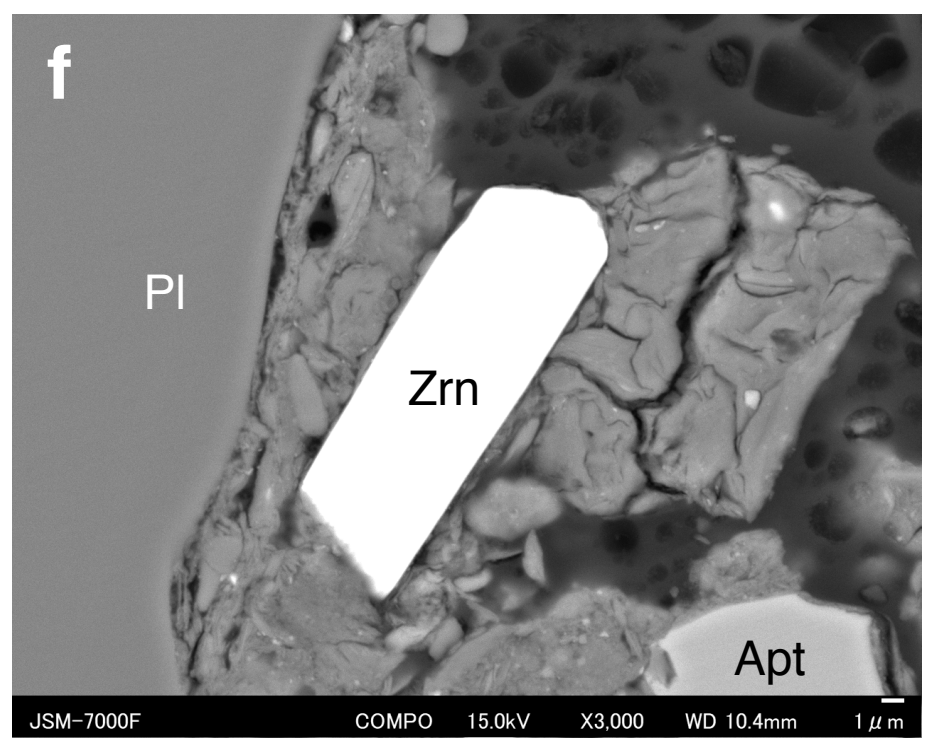

**Figure 1**: Back-scattered electron images showing zirconolite grains (a) Zrl#1, (b) Zrl#2, (c) Zrl#3, (d) Zrl#4, and (e) Zrl #5 and (f) a zircon grain in Erg Chech 002. Pits in Zrl#1 and 2 are the in-situ Pb isotope analysis spots from NanoSIMS. Phase abbreviations are as follow: Apt, apatite; Bd, baddeleyite; Ilm, ilmenite; Mer, merrillite; Mt, metal; Pl, plagioclase; Px, pyroxene; Si, unidentified silica polymorph; Trd, tridymite; Tro, troilite; X, unidentified Ti-Fe-Zr mineral; Y, unidentified Nb-Ti-Zr mineral; Zrl, zirconolite; Zrn, zircon.

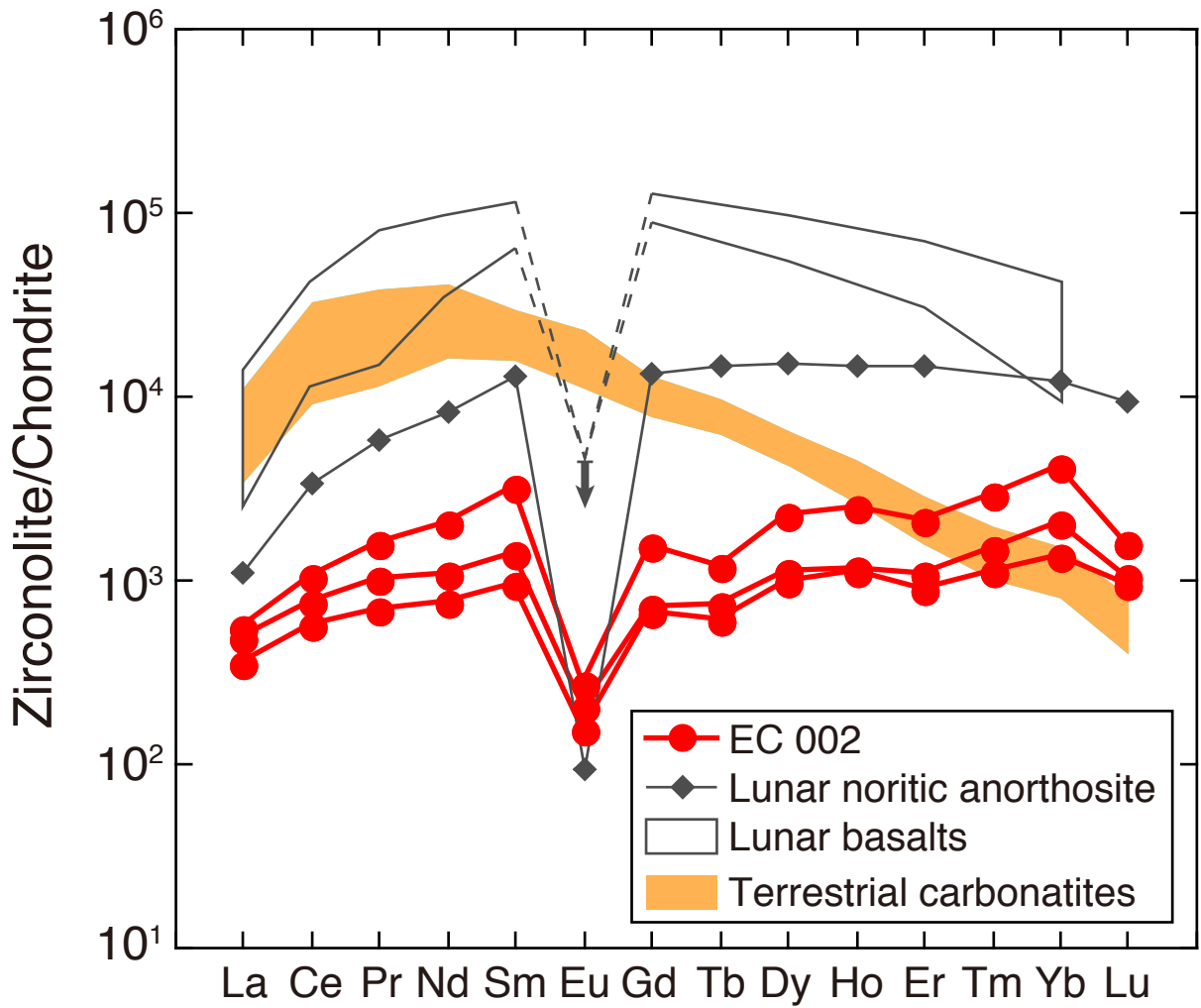

**Figure 2**: Chondrite-normalized REE patterns of zirconolite grains Zrl#1 and 2 in Erg Chech 002. Those in terrestrial carbonatites (Wu et al., 2010), lunar basalts (Rasmussen et al., 2008), and lunar noritic anorthosite (Norman and Nemchin, 2014) are also shown for comparison. The chondrite data are from McDonough and Sun (1995).

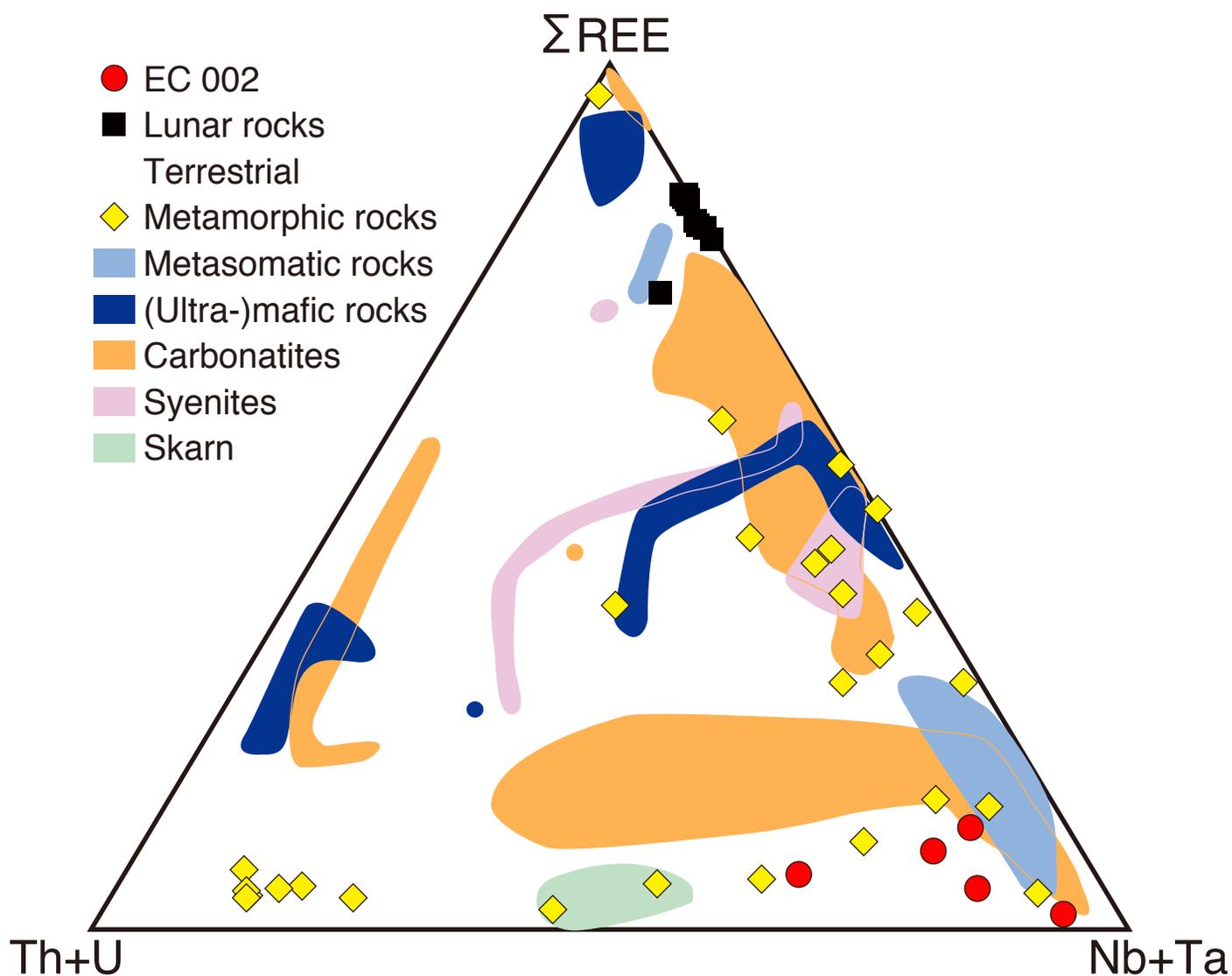

**Figure 3**: Triangular plot of ΣREE–(Th+U)–(Nb+Ta) for the Erg Chech 002 zirconolite. Also shown for comparison are zirconolite compositional fields for lunar rocks and terrestrial syenites, carbonatites, skarn, (ultra-)mafic rocks, metamorphic rocks, and metasomatic rocks. Data and source references are given in Supplementary Table 1.

**Figure 4**: Radiogenic $^{207}$Pb/$^{206}$Pb ratios and the weighted average (bold line) with 95% confidence interval (shaded area). Box heights are 2σ.

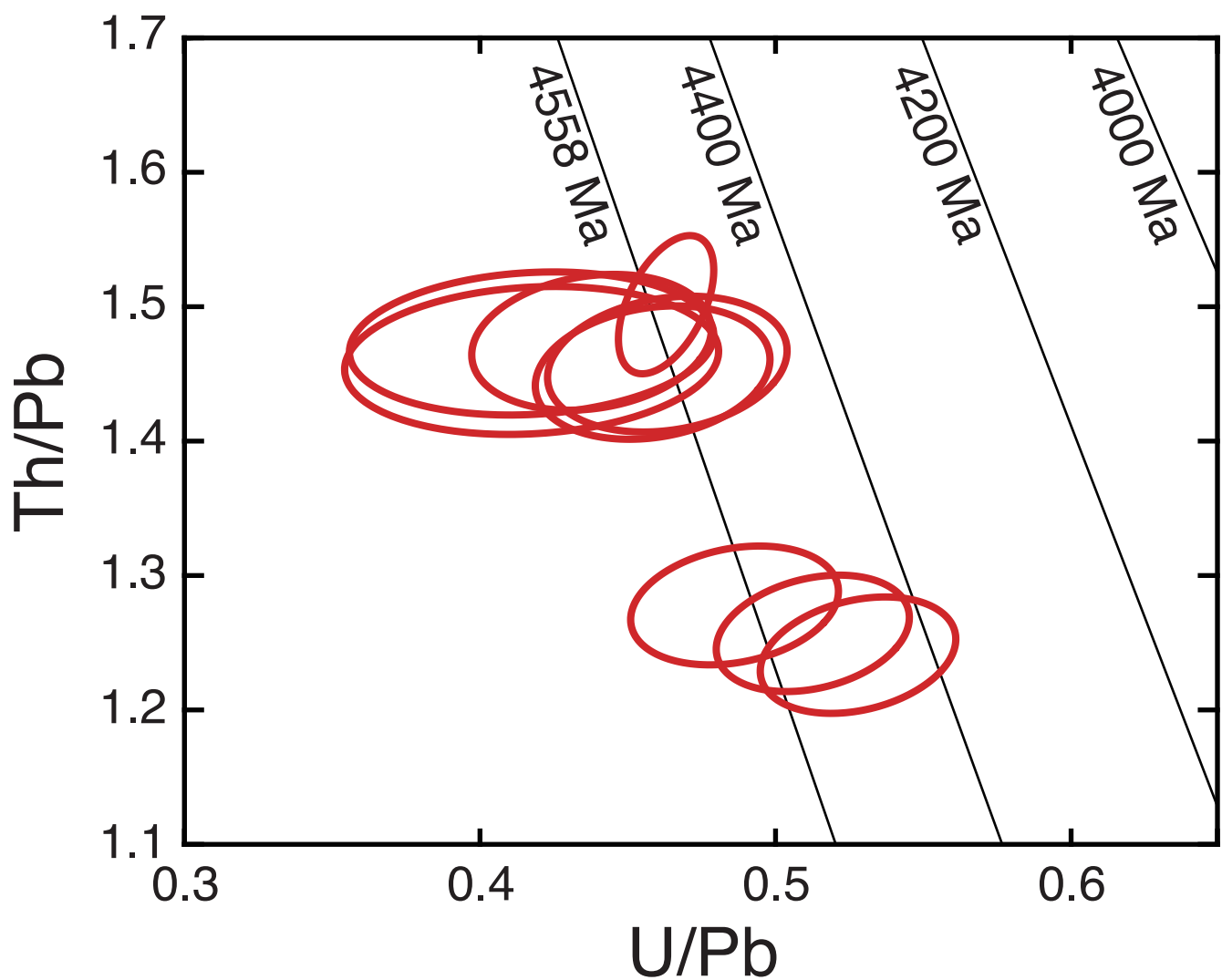

**Figure 5**: Plot of Th/Pb versus U/Pb for zirconolite in Erg Chech 002. Errors at 95% confidence level are represented by ellipses. Chemical reference isochrons of 4558, 4400, 4200, and 4000 Ma are also shown.

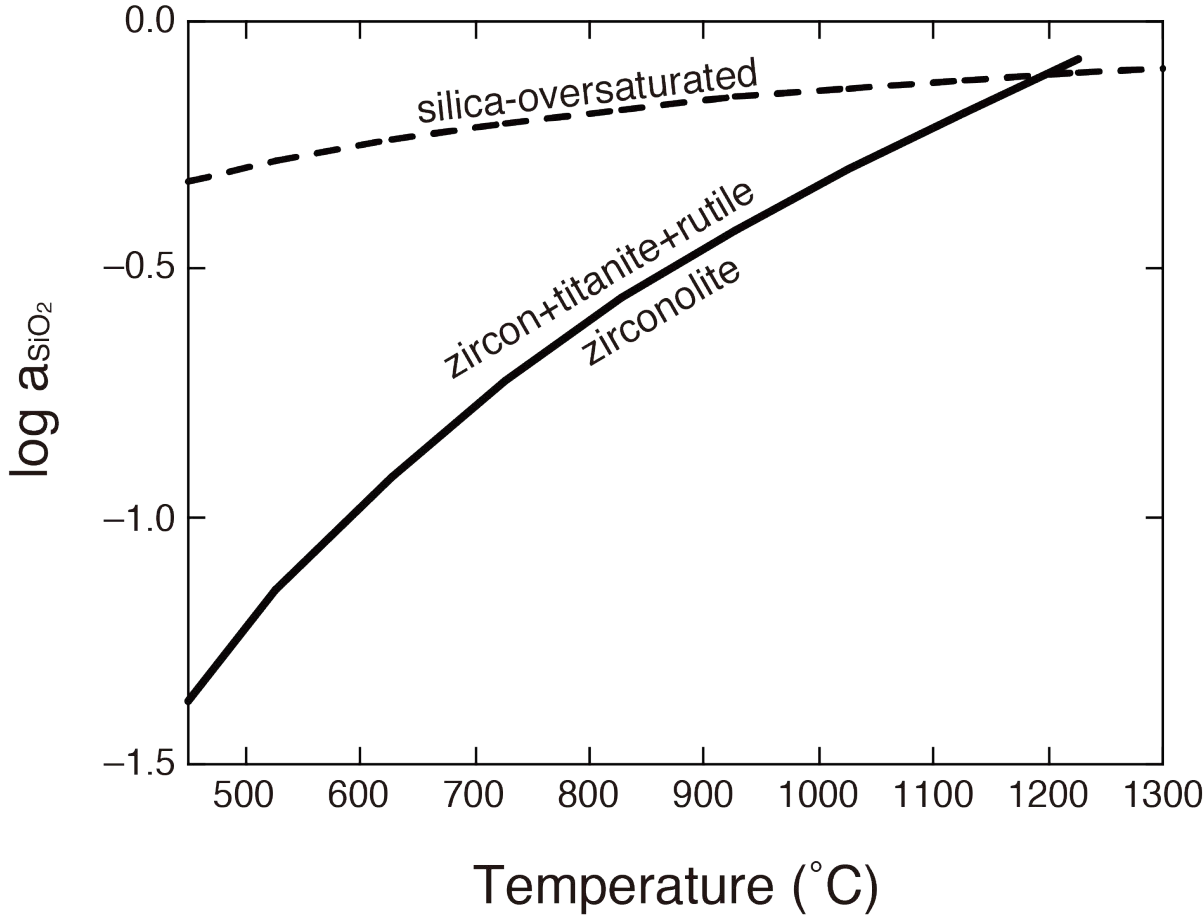

**Figure 6**: Variation of silica activity (aSiO$_2$) with temperature at 1 bar for two silica buffer reactions: SiO$_2$ (glass) ⇔ SiO$_2$ (quartz/cristobalite) and CaZrTi$_2$O$_7$ (zirconolite) + 2SiO$_2$ (glass) ⇔ ZrSiO$_4$ (zircon) + CaTiSiO$_5$ (titanite) + TiO$_2$ (rutile). Thermodynamic data are taken from Putnum et al. (1999) for zirconolite and from Robie and Hemingway (1995) for the other phases.